\documentclass[showpacs]{revtex4}
\usepackage{graphicx}
\usepackage{bm}

\begin{document}

\title{{\Large Conversion of nuclear to 2-flavour quark matter in rotating 
compact stars: A general relativistic perspective }}

\bigskip
\bigskip
\author{\large Ritam Mallick}
\email{ritam@bosemain.boseinst.ac.in}
\affiliation{Department of Physics, Bose Institute, Kolkata - 700009, India}
\author{\large Abhijit Bhattachryya}
\affiliation{Department of Physics, Calcutta University, Kolkata - 700009, India}
\author{\large Sanjay K. Ghosh, Sibaji Raha}
\affiliation{Centre for Astroparticle Physics and Space Science \& Department of Physics, Bose Institute, Kolkata - 700009, India}

\bigskip
\bigskip
\begin{abstract}
\leftskip1.0cm
\rightskip1.0cm
The conversion of neutron star to strange star is argued to be a two 
step process.
The first process involves the deconfinement of nuclear to two-flavour
quark matter.
The GR results shows that the propagating front breaks up into fragments 
which propagate
with different velocities along different directions.
The
time taken for this conversion to happen is of the order of few $ms$.
This calculation indicates the inadequacy of
non-relativistic (NR) or even Special Relativistic (SR) treatments for
these cases.

\end{abstract}

\maketitle

\section{Introduction}
Strange Quark Matter (SQM), consisting of approximately equal numbers of
up ({\it u}), down ({\it d}) and strange ({\it s}) quarks, 
is conjectured to be the 
{\it true} ground state of strong interaction \cite{key-1,key-2}. 
SQM could naturally occur in the cores of compact stars,
where central densities of about an order of magnitude higher than the
nuclear matter saturation density. The transition from 
nuclear (hadronic) to quark matter should proceed
through a conversion to an initial stage of (metastable) two flavour quark
matter, which should decay to the stable SQM. Thus, neutron stars with
sufficiently high central densities ought to get converted to strange,
or at least hybrid, stars. These transitions could have observable
signatures in the form of a jump in the breaking index and gamma ray
bursts  \cite{key-3,key-4}.
\par
There are several plausible scenarios where neutron stars could convert
to quark stars, through a "seed" of external SQM \cite{key-5}, or
triggered by the rise in the central density due to a sudden spin-down
in older neutron stars \cite{key-6}. Several authors have studied the
conversion of nuclear matter to strange matter under different
assumptions \cite{key-7,key-11,key-17}.
\par
In our recent work \cite{key-18}, we have argued that the conversion process
is a two step process. The first process involves the deconfinement of
nuclear to two-flavour quark matter and the conversion process takes some
milliseconds to occur. GR 
effects in such processes is studied by us \cite{key-18a}.
In this article, we make a detailed study of this process.

\section{Theory}
We use the Nonlinear Walecka model for the nuclear matter
euqation of state (EOS). 
We consider the conversion of nuclear
matter, consisting of only nucleons (\textit{i.e.} without hyperons)
to a two-flavour quark matter. The final composition of the quark matter
is determined
from the nuclear matter EOS by enforcing the baryon number conservation
during the conversion process.

The metric describing the structure of the star, is given by \cite{key-20}
\begin{eqnarray}
ds^2 = -e^{\gamma+\rho}dt^2 + e^{2\alpha}(dr^2+r^2d\theta^2) + \nonumber \\
e^{\gamma-\rho}r^2 sin^2\theta(d\phi-\omega dt)^2
\end{eqnarray}
The four gravitational potentials $\alpha, \gamma, \rho$ and $\omega$ are
functions of $\theta$ and $r$ only.
The solution of the star
is obtained from the {\bf 'rns'} code \cite{key-23}.

We heuristically assume the existence of a combustive phase transition
front of infinitesimal thickness, and study the outward propagation of the
front. Let us assume that the conversion front
is generated at the center of the star, and it propagates outwards through
the star with a certain velocity. Employing the conservation 
conditions \cite{key-24} and
further employing the entropy condition \cite{key-26}, we determine the flow
velocity of matter in the two phases $v_1$ and $v_2$.
\par
It is possible to classify the various conversion mechanisms by comparing
the velocities of the respective phases with the corresponding velocities
of sound in these phases. For the conversion to be
physically possible, velocities should satisfy an additional condition,
namely, $0\leq v_{i}^{2}\leq 1$.
If we plot different velocities against
baryon number, we get curves from which we could calculate the initial
velocity of the front at the centre of the star \cite{key-18}.
The results \cite{key-18} also show that the range of values of baryon
density, for which the flow velocities are physical, increases with
temperature.
Starting with this initial velocity, we investigate the evolution 
of the front with time. Treating both
nuclear and strange matters as ideal fluids, the system is governed,
together with the metric and the EOS, by the Einstein's equation
${R_i}^k - \frac{1}{2}{\delta_i}^k R = \kappa {T_i}^k$ and the equation of motion
${T^k}_{i;k}=0$  \cite{key-27}.

\begin{figure}
\vskip 0.4 in
\centering
\includegraphics[width=2.75in]{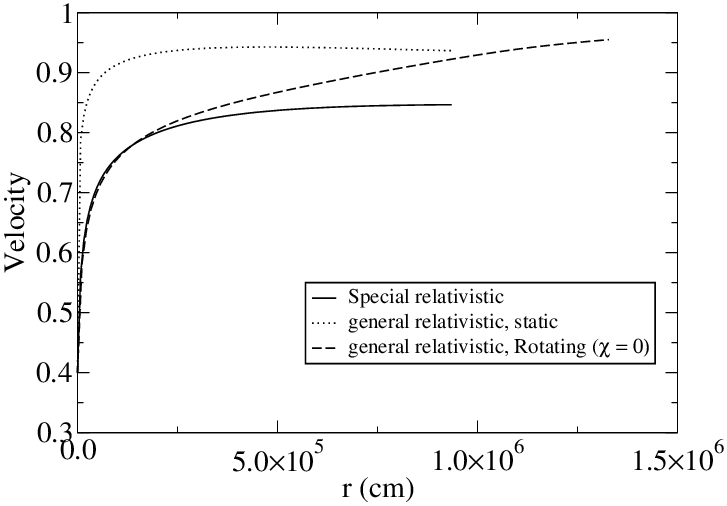}
\caption{Variation of velocity of the front along the radial direction of
 the star for three different cases, namely SR, non-rotating GR and rotating GR.}
\end{figure}

\par
The above two equations are the starting point for deriving the appropriate
continuity and Euler's equations \cite{key-18a}.
After a bit of algebra, we get a single differential equation for $v$:
\begin{equation}
\frac{\partial v}{\partial r} = \frac{W^2v[K+K1]}{2[v^2(1+G)^2-n(1+v^2G)^2]}.
\end{equation}
$\omega=0$ and $sin\theta=1$ in this equation yield the equation for the static star and if we put all potentials equal to zero, we recover our equation for
the SR case \cite{key-18}.

\begin{figure}
\vskip 0.4 in
\centering
\includegraphics[width=2.75in]{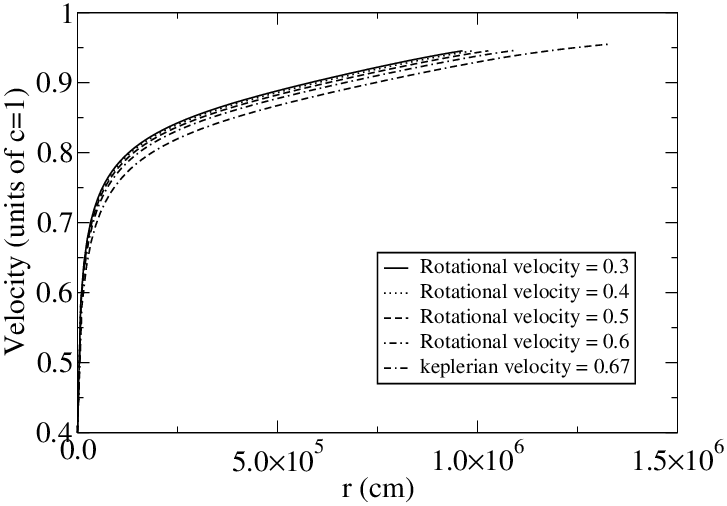}
\caption{Variation of velocity of the front along the radial direction for
 different rotational velocity of the star.}
\end{figure}

\section{results and discussion}
Having constructed the density profile of
the star for a fixed central density, the respective
flow velocities $v_{1}$ and $v_{2}$ of the nuclear and quark matter in the
rest frame of the front, at a radius infinitesimally close to the center of
the star. This would give us the initial velocity of the front ($-v_{1}$),
at that radius, in the nuclear matter rest frame. With this, we integrate
eqn. (2) outwards along the radius of the star. The solution gives the
variation of the velocity with position as a function of the time of arrival
of the front. Using this velocity profile, we can calculate the time required
to convert the whole star using the relation $\frac{dr}{d\tau}=vG$.

\begin{figure}
\vskip 0.4 in
\centering
\includegraphics[width=2.75in]{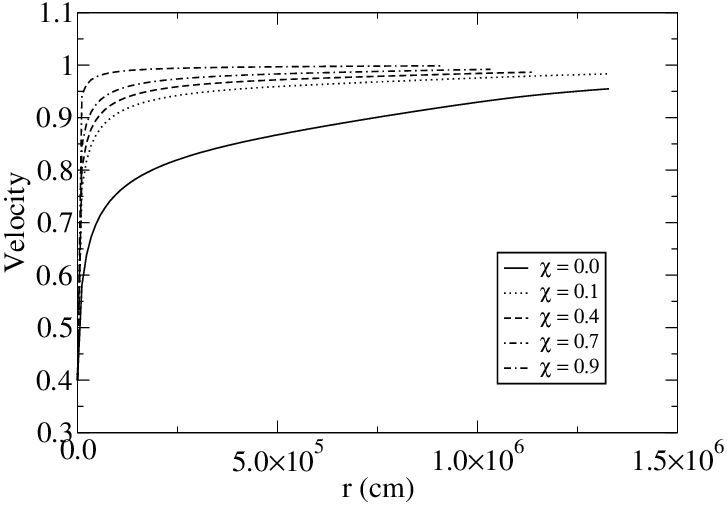}
\caption{Variation of velocity of the front along the radial direction for
 different  $\chi$.}
\end{figure}

For a rotating star, due to the asymmetry, we introduce a new parameter 
$\chi=cos\theta$, along the
vertical axis of the star. We start our calculation by choosing the central
density of the star to be $7$ times the nuclear matter density, for which the
Keplerian velocity of the star is $0.67 \times 10^{-4} sec^{-1}$ (the
rotational velocities given in fig. (2) are all in units of $10^{-4} sec^{-1}$).
For this central density, the initial velocity of the front comes out to
be $0.45  $. In fig. (1) the propagation velocity
of the front along the radial direction of the star for three cases. 
The unbroken curve is for the
SR case, the broken curve for non-rotating GR case and the dotted curve for
the rotating GR case with $\chi=0$, {\it i.e} at the equator. 
Due to the asymptotic behaviour the velocity shoots up at the centre 
and saturates at larger radii.
It can also be clearly
seen that the GR effect increases the velocity of the front considerably
(maximum by $30 \%$) and the effect is most pronounced for the static case.
The rotational effect of the star seems to suppress the GR effect and
therefore the velocity of the front decreases. The result becomes clearer if
we look at fig. (2) where we have plotted the front velocity with equatorial
radius for different rotational velocities; as the rotational velocity
increases, the velocity of the front decreases.
\begin{figure}
\vskip 0.4 in
\includegraphics[width=2.75in]{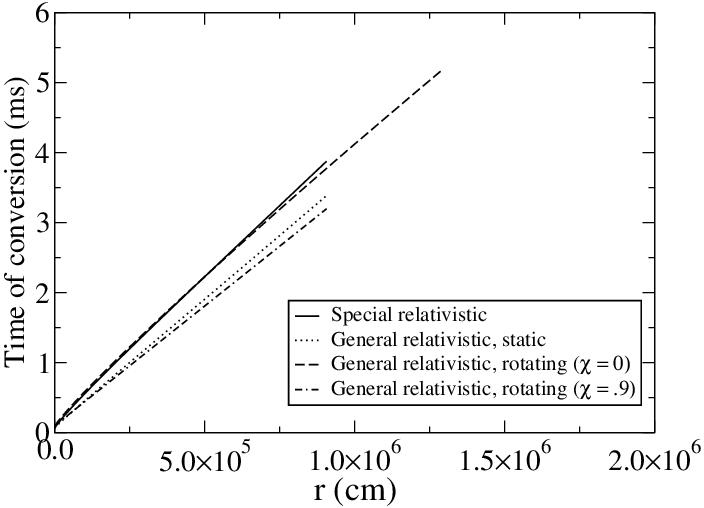}
\caption{Variation of time of arrival of the front at certain radial distance
 for different cases.}
\end{figure}

From fig. (3), we find that the front velocity is maximum along the polar
direction and minimum along the equator. Therefore, at any particular 
instant of time, we may have a
situation where the polar part of the star has been converted while along the
equatorial direction, the front is still propagating.
\par
From fig. (4) we can see that the time taken by the conversion front to
convert the neutron star to two-flavour quark star is of the order of few $ms$. The static star takes the minimum time ($3.3 ms$) whereas the rotating star
takes the maximum time ($5.1 ms$) due to the enlarged equatorial radius. The
polar part of the star needs much lesser time for conversion ($3.1 ms$),
even less than static star, as its radius gets compressed. 

\par
To summarize, we have shown in this article that the conversion of nuclear
matter to quark matter in compact stars, especially rotating stars which
are more realistic than static stars, is strongly affected by GR effects.
The emergence of different conversion fronts, propagating with different
velocities along different radial directions was not
anticipated by Newtonian or SR calculations. It remains to be explored whether
the incorporation of dissipative effects materially changes the results.
Though the calculation is much involved, such an
investigation is on our immediate agenda.
\par

\end{document}